\documentclass[sigconf, nonacm]{acmart}
\AtBeginDocument{%
  }

\setcopyright{acmlicensed}
\copyrightyear{2025}
\acmYear{2025}
\acmDOI{XXXXXXX.XXXXXXX}
\acmConference[Conference acronym CIKM 2025]{Advances in Financial AI:
Innovations, Risk, and Responsibility
in the Era of LLMs}{November 14th, 2025}{COEX in Gangnam, Seoul, South Korea}
\acmISBN{978-1-4503-XXXX-X/2018/06}

\usepackage{listings}
\usepackage[most]{tcolorbox} 

\usepackage{graphicx} 
\begin{document}

\title{FINCH: Financial Intelligence using Natural language for Contextualized SQL Handling}

\author{Avinash Kumar Singh}
\affiliation{%
  \institution{Domyn}
  \city{Hyderabad}
  \country{India}
}
\email{avinash.kumarsingh@domyn.com}

\author{Bhaskarjit Sarmah}
\affiliation{%
  \institution{Domyn}
  \city{Gurgaon}
  \country{India}
}
\email{bhaskarjit.sarmah@domyn.com}

\author{Stefano Pasquali}
\affiliation{%
  \institution{Domyn}
  \city{New York}
  \country{USA}
}
\email{stefano.pasquali@domyn.com}

\begin{abstract}
  Text-to-SQL, the task of translating natural language questions into SQL queries, has long been a central challenge in NLP. While progress has been significant, applying it to the financial domain remains especially difficult due to complex schema, domain-specific terminology, and high stakes of error. Despite this, there is no dedicated large-scale financial dataset to advance research, creating a critical gap. To address this, we introduce a curated financial dataset (FINCH) comprising 292 tables and 75,725 natural language–SQL pairs, enabling both fine-tuning and rigorous evaluation. Building on this resource, we benchmark reasoning models and language models of varying scales, providing a systematic analysis of their strengths and limitations in financial Text-to-SQL tasks. Finally, we propose a finance-oriented evaluation metric (FINCH Score) that captures nuances overlooked by existing measures, offering a more faithful assessment of model performance.
\end{abstract}


\maketitle

\section{Introduction}
\par The task of translating natural language questions into SQL queries, commonly known as Text-to-SQL, has remained a central research challenge since the early days of relational databases\cite{wong2021survey}. Early approaches were dominated by rule-based grammars and symbolic parsing, but the advent of neural methods fundamentally reshaped the field. The introduction of Seq2SQL (2017) \cite{Zhong2017Seq2SQLGS} showed how reinforcement learning could be used to generate executable queries that surpassed template-based systems. Soon after, SQLNet \cite{Xu2017SQLNetGS} replaced reinforcement learning with a sketch-based decoding strategy, reducing complexity while maintaining strong performance. The release of Spider (2018) \cite{Yu2018SpiderAL}, the first large-scale cross-domain dataset, provided a foundation for evaluating generalization across diverse schemas and quickly became the benchmark of choice. Follow-up datasets such as CoSQL and SParC (2019) \cite{Yu2019CoSQLAC, Yu2019SParCCS} extended the task into conversational and context-dependent settings. Progress at the model level soon followed, with RAT-SQL (2020) \cite{Wang2019RATSQLRS} introducing relation-aware transformers for schema linking and PICARD (2021) \cite{Scholak2021PICARDPI} enforcing constrained decoding for large pre-trained language models. More recently, the BIRD benchmark (2023) \cite{Li2023CanLA} scaled evaluation to billions of queries, creating the most comprehensive resource to date for measuring robustness at scale.

\par Despite this trajectory of progress, significant challenges remain. Many state-of-the-art systems achieve high benchmark performance yet falter on schemas that diverge from their training corpus \cite{Shi2021Ada, Lei2020Tracing}. While Spider \cite{Yu2018SpiderAL} and BIRD \cite{Li2023CanLA} emphasize cross-domain generalization, their schemas mostly reflect encyclopedic or open-domain knowledge, rather than industry-specific settings. Schema linking methods such as RAT-SQL \cite{Wang2019RATSQLRS} and DCG-SQL (2025) \cite{Wang2025DCGSQL} improve contextual alignment, but remain sensitive to noisy descriptions and wide column spaces—conditions common in enterprise databases \cite{Gan2021Natural, Pourreza2023DINSQL, Scholak2021PICARDPI}. More recent approaches, such as NL2PY2SQL (2024) \cite{Liu2024NL2PY2SQL}, explore intermediate Python-to-SQL representations, yet questions remain about their efficiency and interpretability in high-stakes domains. Finally, evaluation methodologies continue to rely on exact match \cite{Yu2018SpiderAL} and execution accuracy \cite{Zhong2017Seq2SQLGS}, metrics that often fail to capture subtle but meaningful structural differences. These limitations collectively expose a gap between benchmark-driven advances and the demands of real-world deployment.

\par In parallel, the financial domain has witnessed growing interest in executable reasoning tasks that, while not always SQL-focused, address challenges directly relevant to database querying \cite{Lan2022UniRPG, Qin2021NeuralSymbolic}. FinQA (2021) \cite{Chen2021FinQAAD} introduced a benchmark for numerical reasoning over corporate financial reports, pairing textual inputs with program-of-operations supervision to foster neural–symbolic integration. Similarly, TAT-QA (2021) \cite{Zhu2021TATQAAQ} combined tabular and textual information, reflecting the multimodal nature of financial data. ConvFinQA (2022) \cite{Chen2022ConvFinQAET} further extended this into multi-turn reasoning, simulating analyst workflows where queries evolve iteratively. Beyond datasets, modular approaches have emerged. For example, Nye et al. \cite{Nye2021Structuring} proposed a system where one agent extracts key performance indicators from filings while another translates them into SQL queries, distributing reasoning across specialized components. Alongside these advances, evaluation metrics have shifted from rigid exact match to tree-edit distances \cite{Zhang2019Editing} and structure-aware similarity measures \cite{Zhou2022Evaluating}, more aligned with analyst expectations.

\par Recent efforts have begun to address Text-to-SQL specifically in finance, though the field is still nascent compared to open-domain research. FinSQL (2024) \cite{Zhang2024FinSQLML} made the first dedicated attempt by releasing the BULL dataset, consisting of wide-schema financial tables designed for cross-database generalization. Its results showed that domain-specific tuning could yield up to 36\% relative improvement over general-purpose baselines, but also highlighted steep performance drops on larger schemas. BookSQL (2024) \cite{Kumar2024BookSQLAL} complemented this by offering 75k natural language–SQL pairs in the accounting domain, exposing the limitations of large pre-trained models when handling subtle business semantics and complex joins. On the methodological side, NL2PY2SQL (2024) \cite{Gupta2024NL2PY2SQL} introduced a two-step natural language–Python–SQL pipeline that better captured domain logic, while DCG-SQL (2025) \cite{Li2025DCGSQL} leveraged deep contextual graph representations to enhance schema linking in wide and relationally complex tables. Beyond accuracy, Reliable Text-to-SQL (RTS, 2025) \cite{Wang2025RTS} incorporated adaptive abstention and human-in-the-loop corrections, highlighting reliability as a core requirement in banking, auditing, and investment contexts.

\par Taken together, these developments highlight a critical gap: despite steady improvements in benchmark performance, current systems remain ill-suited for financial deployment \cite{Shen2023Enterprise, Deng2024FinTextSQL, Yang2025TrustSQL}. Financial databases present unique challenges—wide and evolving schemas, strict domain constraints, regulatory compliance, and heightened sensitivity to errors. Addressing these challenges requires domain-specific datasets, architectures explicitly tailored to financial schemas, and evaluation methodologies that extend beyond exact match to capture semantic fidelity and operational safety. Bridging this gap is not merely an incremental step from general Text-to-SQL research, but a necessary shift toward building practical systems capable of supporting analysts, auditors, and policymakers in critical decision-making.

\par While financial QA and domain-specific Text-to-SQL have advanced considerably, most prior efforts remain centered on document centric or hybrid table–text reasoning rather than direct querying over financial SQL databases. As a result, dedicated Text-to-SQL benchmarks for finance are scarce, and the community still lacks a resource that faithfully captures the intricacy, terminology, and precision demanded by real-world financial applications. Addressing this gap requires both large-scale, domain-specific datasets and evaluation methodologies that reflect the unique challenges of financial environments.  

\par Motivated by these challenges, this work makes three key contributions to advance Text-to-SQL in the financial domain:
\begin{itemize}
\item \textbf{Large-scale financial dataset curation:} We consolidate and extend existing resources such as BIRD \cite{Li2023CanLA}, Spider \cite{Yu2018SpiderAL}, FinSQL \cite{Zhang2024FinSQLML}, and BookSQL \cite{Kumar2024BookSQLAL} into a unified dataset. All SQL queries are normalized for compatibility with SQLite, ensuring broad usability within the open-source community. The final dataset (FINCH) spans 33 databases across domains including retail, banking transactions, loans, insurance, sales, marketing, finance, e-commerce, funds, stocks, and accounting. In total, it contains 292 tables, 2233 columns, 177 relations, and 75,725 NL–SQL pairs—providing the largest finance-oriented benchmark, suitable for both evaluation and fine-tuning.  

\item \textbf{Comprehensive model benchmarking:} We evaluate four categories of models: large-scale LLMs (Qwen3-235B-A22B\footnote{https://huggingface.co/Qwen/Qwen3-235B-A22B}), medium- and small-scale open-source models (GPT-OSS-120B\footnote{https://huggingface.co/openai/gpt-oss-120b}, GPT-OSS-20B\footnote{https://huggingface.co/openai/gpt-oss-20b}, Qwen3-8B\footnote{https://huggingface.co/Qwen/Qwen3-8B}), and reasoning-centric models (Phi-4-mini-reasoning\footnote{https://huggingface.co/microsoft/Phi-4-mini-flash-reasoning} and Arctic-Text2SQL-R1-7B\footnote{https://huggingface.co/Snowflake/Arctic-Text2SQL-R1-7B}). Results show that GPT-OSS-120B achieves the strongest overall performance, surpassing even larger models like Qwen3-235B-A22B, while Arctic-Text2SQL-R1-7B—despite its modest parameter scale—emerges as the third-best performer. These findings highlight that reasoning-centric architectures, when carefully adapted, can rival and even outperform much larger general-purpose LLMs in financial Text-to-SQL tasks.  

\item \textbf{Finance-specific evaluation metric:} We propose a tailored evaluation metric (FINCH Score) that integrates component matching and execution accuracy with weighted scoring and tolerance thresholds. This design alleviates disproportionate penalties for minor floating-point mismatches while emphasizing structural fidelity—prioritizing correctness of column names, table references, and conditions. Experiments demonstrate that this metric provides a more faithful assessment of performance in financial settings, better aligning with practical requirements than traditional exact-match measures.  
\end{itemize}

\section{FINCH Dataset Construction}
To address the pressing need for a large-scale, finance-oriented Text-to-SQL benchmark, 
we curated the \textbf{FINCH} dataset by thoughtfully consolidating and refining four publicly 
available resources: \textit{BIRD} \cite{Li2023CanLA}, \textit{Spider} \cite{Yu2018SpiderAL}, 
\textit{BULL} \cite{Zhang2024FinSQLML}, and \textit{BookSQL} \cite{Kumar2024BookSQLAL}.  

Although \textit{Spider} and \textit{BIRD} are widely recognized benchmarks in the Text-to-SQL 
community, they were not originally designed with financial applications in mind. Their databases 
span a broad range of domains---from sports and education to entertainment and geography---many 
of which have little relevance to financial systems. In contrast, practitioners and researchers 
working with financial data face highly specialized challenges, such as interpreting 
banking records, analyzing card transactions, managing retail and sales data, or handling 
loan and insurance queries \cite{Chen2021FinQAAD, Zhu2021TATQAAQ}. Using generic benchmarks 
risks diluting the focus of models and limits their applicability in real-world financial settings.  

To bridge this gap, we undertook a careful screening process to retain only those databases that 
align with financial contexts. Specifically, we identified and preserved domains covering retail 
sales, card transactions, banking services, loan processing, insurance, and e-commerce. By doing 
so, we ensured that FINCH is not merely another aggregation of Text-to-SQL data, but a benchmark 
with direct utility for finance-related applications.  

Table~\ref{tab:dataset_comparison} compares FINCH with four widely used Text-to-SQL datasets. 
Unlike prior benchmarks that either emphasize scale (e.g., BookSQL) or cross-domain generalization 
(e.g., Spider, BIRD), FINCH balances scale, schema diversity, and financial relevance. In particular, 
it offers 33 finance-specific databases with a higher average table-to-database ratio than earlier 
resources, making it especially suited for evaluating schema linking and compositional reasoning in 
finance.  

\begin{table}[ht!]
\caption{Comparison of FINCH with existing Text-to-SQL datasets. Columns indicate dataset size (\#Size), number of databases (\#DB), and (\#T/DB) database-to-table ratio.}
\centering
\label{tab:dataset_comparison}
\begin{tabular}{lccccc}
\toprule
\textbf{Dataset} & \textbf{\#Size} & \textbf{\#DB} & \textbf{\#DB*} & \textbf{\#Size*} & \textbf{\#T/DB} \\
\midrule
Spider   & 10,181  & 200   & 22  & 1,100  & 5.1 \\
BIRD     & 12,751  & 95    & 7   & 812    & 7.3 \\
BULL     & 4,966   & 3     & 3   & 4,966  & 26.0 \\
BookSQL  & 78,433  & 1     & 1   & 68,907 & 7.0 \\
\hline
\textbf{FINCH} & \textbf{-} & \textbf{-} & \textbf{33} & \textbf{75,725} & \textbf{8.85} \\
\bottomrule
\end{tabular}
\end{table}

\subsection{Dataset Curation}
The curation of FINCH was carried out in several deliberate and methodical steps. From 
\textit{Spider}, we selected 22 databases that carried clear financial relevance, and from 
\textit{BIRD} we retained 7 such databases. For \textit{BULL} and \textit{BookSQL}, which are 
already finance-focused by design, we preserved all available samples where both natural language 
questions and their corresponding SQL statements were present. This resulted in a diverse yet 
domain-specific dataset, structured as follows:

\begin{itemize}
    \item \textbf{BIRD}: 1,139 samples
    \item \textbf{Spider}: 1,100 samples
    \item \textbf{BULL}: 4,966 samples
    \item \textbf{BookSQL}: 78,433 samples
\end{itemize}

During this process, we observed that the \textit{BookSQL} test split lacked corresponding SQL 
queries, and therefore excluded this portion from FINCH to preserve completeness and usability.  
To further ensure data integrity, every SQL query was executed against its associated SQLite 
database. This validation step proved crucial, revealing a substantial number of anomalies that, 
if left unaddressed, would have undermined the reliability of the dataset. Specifically, we found:

\begin{itemize}
    \item In \textit{BIRD}, 327 out of 1,139 queries failed, including 198 with incorrect 
    column names and 129 with invalid table references.
    \item In \textit{BULL}, 60 out of 4,966 queries were problematic, with 43 syntax errors, 
    10 wrong table references, 1 column error, 4 incomplete queries, and 2 unrecognized tokens.
    \item In \textit{BookSQL}, 9,526 out of 78,433 queries exhibited errors: 7,018 incorrect 
    column names, 1,494 invalid table references, and 1,014 syntax errors.
    \item Interestingly, the \textit{Spider} subset stood out with no anomalies detected.
\end{itemize}

The process of identifying, cataloguing, and correcting these errors was not purely mechanical. 
It required repeated query execution, close inspection of database schemas, and contextual reasoning 
about the intent behind each question. This diligence ensured that FINCH not only aggregates data 
at scale but also maintains the precision and robustness necessary for finance-specific 
Text-to-SQL research.  

The final version of FINCH consists of \textbf{33 databases}, encompassing \textbf{292 tables}, 
\textbf{2,233 columns}, and \textbf{177 relations}, with a total of over \textbf{75,725 NL–SQL pairs}. 
In terms of difficulty distribution, FINCH includes 9,358 easy, 33,780 medium, and 32,587 hard examples. 
An additional 7,035 samples that were originally unlabeled were annotated following the schema complexity 
guidelines proposed in \cite{Kumar2024BookSQLAL}. This design ensures coverage across diverse financial 
operations, schema complexities, and SQL constructs. Furthermore, FINCH makes extensive use of 
SQL operations such as \texttt{ORDER BY}, \texttt{GROUP BY}, and nested queries, which are essential 
for modeling complex reasoning in financial environments.  

\begin{figure*}[ht]
\centering
\includegraphics[width=0.9\textwidth]{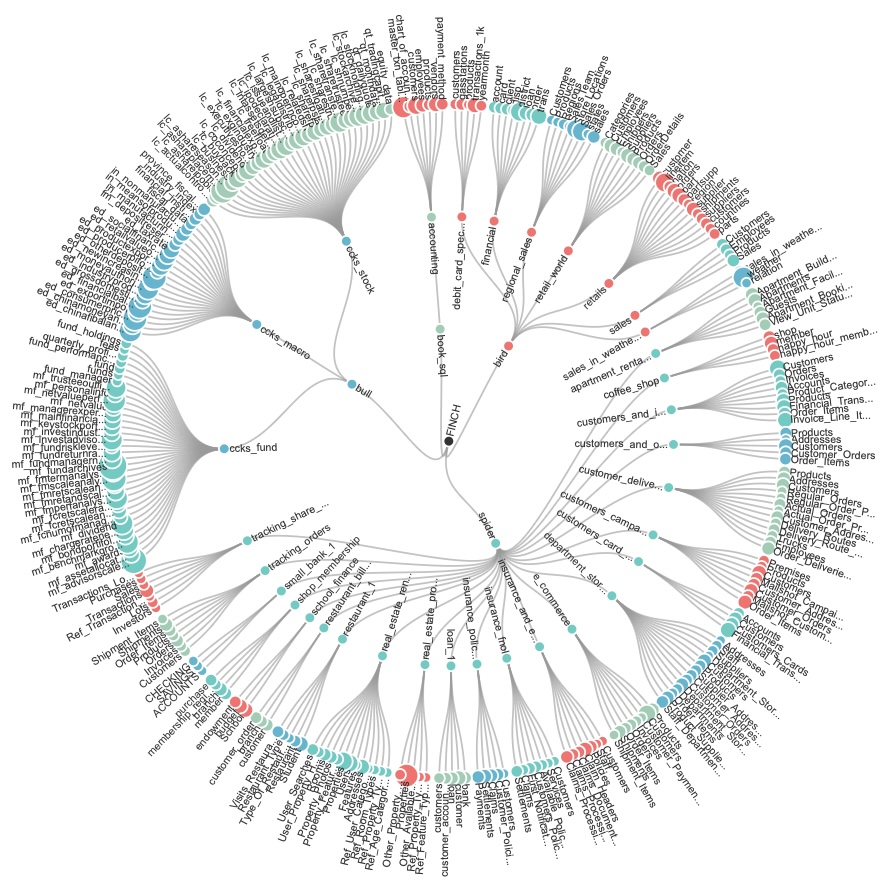}
\caption{Representation of the FINCH dataset showing the integration of different databases and tables across financial domains.}
\label{fig:dataset_information}
\end{figure*}

Figure~\ref{fig:dataset_information} provides a high-level overview of FINCH, illustrating how 
databases from multiple sources were combined into a single, unified benchmark. The figure also 
highlights the diversity of database schemas and the breadth of domain coverage achieved through 
this consolidation.  

In summary, FINCH represents the first large-scale, finance-specific Text-to-SQL dataset that combines 
breadth (multiple domains), depth (rich schema complexity), and difficulty (non-trivial SQL constructs). 
We believe FINCH will serve as a cornerstone resource for evaluating and fine-tuning both large language 
models and reasoning-centric architectures in financial applications.

\section{Experiment Setup and Evaluation Metrics}
\subsection{Experimental Setup}
We benchmark a diverse set of models, with particular emphasis on those explicitly optimized for \emph{reasoning}, to test whether such capabilities transfer effectively to Text-to-SQL in the financial domain. Concretely, our evaluation spans: (i) \textbf{large-scale language models} such as Qwen3-235B-A22B and GPT-OSS-120B, (ii) \textbf{medium and small-scale models} such as Qwen3-8B and GPT-OSS-20B, and (iii) \textbf{reasoning-centric models} including \textit{Phi-4-mini-reasoning} and \textit{Arctic-Text2SQL-R1-7B}, which serve as the strongest reasoning-oriented baselines in our study.  

This selection is motivated by three factors. First, the \textbf{scaling contrast}—ranging from models with hundreds of billions to those with single-digit billions of parameters—allows us to examine whether SQL structural fidelity improves predictably with model size, as suggested by scaling law studies \cite{Kaplan2020ScalingLF, Hoffmann2022TrainingCL}. Second, the inclusion of models from different design families (\texttt{Qwen}-like, \texttt{GPT}-like, and \texttt{Arctic}) ensures \textbf{family diversity}, thereby mitigating biases associated with a single architectural lineage. Finally, the \textbf{reasoning alignment} dimension reflects a growing emphasis on models that advertise enhanced chain-of-thought reasoning and tool-use capabilities \cite{Wei2022ChainOT, Kojima2022LargeLM, Yao2023ReActSY}, which we hypothesize will improve schema grounding, operator selection, and compositional joins—capabilities that are especially critical in financial Text-to-SQL tasks.  

To ensure comparability across models, we adopt a uniform one-shot prompting protocol (see Table~\ref{tab:prompt_example}). Each prompt consists of a natural-language question paired with the corresponding database schema, and models are required to generate a single, syntactically valid, schema-faithful SQL query without additional commentary. The template, stop sequences, and post-processing are held constant: outputs are stripped of code fences, restricted to SQL-only responses, and parsed for validity. This design ensures that observed performance differences can be attributed to model behavior rather than to prompt engineering or decoding artifacts.
The prompt template enforces a controlled setup that guarantees consistency across all evaluated systems. It clearly specifies the task, provides both the natural language input and schema context, and imposes strict constraints on query generation—such as schema fidelity, syntactic correctness, and avoidance of unsupported assumptions. By standardizing these instructions, we minimize variability due to prompting choices and directly test the intrinsic ability of models to translate financial questions into executable SQL.
This level of control is particularly important in the financial domain, where even small deviations—such as an unwarranted join, a misnamed column, or an ill-formed aggregation—can lead to materially incorrect results. Furthermore, the constraints of the prompt are tightly aligned with our evaluation metrics: syntactic correctness underpins execution accuracy \cite{Zhong2017Seq2SQLGS}, schema fidelity is captured by component-matching metrics \cite{Yu2018SpiderAL}, and logical soundness corresponds to structure-aware evaluation measures \cite{Zhou2022Evaluating}. Taken together, the prompt design and evaluation criteria create a tightly coupled FINCH Score that provides a faithful assessment of model performance in realistic financial Text-to-SQL scenarios.

\begin{table}[ht]
\caption{Example one-shot prompt used in all model evaluations. The prompt specifies the task, presents the natural language question with the database schema, and enforces strict SQL generation constraints.}
\label{tab:prompt_example}
\begin{tcolorbox}
\begin{lstlisting}
You are an expert SQL assistant. Your task is to understand Question, have a look at the database Schema and then help the user to find a syntactically correct and optimized SQL query.

### Task:
Generate an SQL query based on the user question and the schema.

### Natural Language Question:
{QUESTION}

### Database Schema:
{DATABASE}

### Constraints:
- Use standard SQL syntax.
- Only use the tables and columns specified in the Database schema.
- Avoid any assumptions or joins not evident in the schema.
- If aggregation or filtering is needed, ensure logical correctness.

### Expected Output:
```sql
\end{lstlisting}
\end{tcolorbox}
\end{table}

\subsection{Evaluation Metrics}
Evaluation of Text-to-SQL systems typically relies on a few standard metrics. \textbf{Exact Matching (EM)}\cite{Yu2018SpiderAL} measures whether the predicted SQL query matches the gold query string-for-string. \textbf{Execution Accuracy (EX)}\cite{Zhong2017Seq2SQLGS} considers a query correct if its output matches the gold query output, treating correctness as binary (1/0). \textbf{Component Matching (CM)}\cite{Yu2018SpiderAL} compares queries at the clause level, such as SELECT, WHERE, GROUP BY, and others, to give a finer-grained picture of correctness. Finally, the \textbf{Valid Efficiency Score (VES)}\cite{Li2023CanLA} extends these ideas by rewarding queries that are not only correct but also efficient to execute. Together, these metrics provide surface-level assessments of query accuracy, execution faithfulness, and efficiency.  

\par However, these metrics fall short in capturing what truly matters for finance. EM and EX are overly strict: they assign zero credit for near-miss queries where differences are purely cosmetic, such as aliasing or harmless reordering. EX is also \emph{materiality-blind}, penalizing queries with small rounding errors or negligible deviations that carry no financial significance. CM assumes all clauses contribute equally, whereas in finance the WHERE, JOIN, GROUP BY, HAVING, and AGG clauses carry the most semantic weight. VES, while useful for measuring efficiency, may unfairly penalize queries that are legitimately complex due to rollups, risk decompositions, or compliance-driven joins. As a result, these standard metrics can over-penalize immaterial differences and fail to capture the clauses and tolerances that encode the true financial meaning of SQL.  

\par To overcome these shortcomings, we propose the FINCH metric, which integrates clause-sensitive structural scoring with execution accuracy under tolerance. FINCH introduces interpretable controls to balance structure and execution, aligns with financial principles of materiality, and remains compatible with existing benchmarks, offering a more faithful evaluation of financial Text-to-SQL performance.

\subsubsection*{Proposed Metric: FINCH Score}

Let the gold SQL be $q^*$ and the model SQL be $\hat{q}$.

\subsubsection*{1) Component-wise Score (Structure/Semantics)}
Choose a component set $K$ (e.g., SELECT, WHERE, GROUP BY, HAVING, ORDER BY, JOIN, AGG, LIMIT, SUBQUERY).  
For each component $k \in K$, compute a similarity
\[
s_k(\hat{q}, q^*) \in [0,1]
\]
(e.g., exact match, set-F1, or token-F1). Assign a weight $w_k \ge 0$ with $\sum_{k \in K} w_k = 1$. The weighted component score is
\begin{equation}
S(\hat{q}, q^*) = \sum_{k \in K} w_k \, s_k(\hat{q}, q^*) \in [0,1].
\end{equation}

\par In the financial context, it is natural to place higher weight on WHERE, JOIN, GROUP BY, HAVING, and AGG, as these clauses encode business rules, portfolio rollups, and compliance filters. For example, in a query computing Value-at-Risk (VaR) for a set of portfolios, omitting the WHERE clause that filters for “active positions” is far more damaging than missing an ORDER BY clause. Weights $w_k$ can be estimated empirically by analyzing historical misqueries: components whose misinterpretations led to significant financial deviations (e.g., millions of dollars in exposure misreported) should be weighted higher than those with negligible impact. This aligns the metric with real-world financial stakes.

\subsubsection*{2) Execution Accuracy (with Tolerance)}
Define execution similarity $e(\hat{q}, q^*) \in \{0,1\}$ with tolerance $\tau$:
\begin{equation}
e(\hat{q}, q^*) =
\begin{cases}
1, & \text{if } \dfrac{|\,r_{\hat{q}} - r_{q^*}\,|}{\max\{1, |r_{q^*}|\}} \leq \tau, \\[10pt]
0, & \text{otherwise}.
\end{cases}
\end{equation}
where $r_{\hat{q}}$ and $r_{q^*}$ are query results and $\tau$ (e.g., $10^{-4}$ or $0.01\%$) enforces a materiality-aware tolerance.

\par This tolerance-based evaluation is essential for finance. Consider a query that computes the average net asset value (NAV) of a mutual fund portfolio. If the predicted query differs by 0.0001 due to rounding or floating-point precision, EM and EX would mark it entirely wrong, even though no financial decision would be affected. By incorporating $\tau$, FINCH respects the principle of materiality, which is central in accounting and regulatory standards \cite{ifrs2018conceptual}. Tolerances can be set in line with domain practices: for daily portfolio valuations, $\tau$ may be very small (e.g., $10^{-5}$), while for aggregated risk rollups, higher tolerances (e.g., $0.1\%$) may be acceptable.

\subsubsection*{3) Combined Metric}
Balance structure and execution via a penalized multiplicative envelope:
\begin{equation}
\text{Score}(\hat{q}, q^*) =
S(\hat{q}, q^*)^{\beta} \cdot \big( \delta + (1-\delta) \, e(\hat{q}, q^*) \big)
\end{equation}
with $\beta \ge 1$ controlling structural fidelity and $\delta \in [0,1)$ the harshness of execution failure.  
Strict: $\delta = 0$; Finance-friendly: $\delta \in [0.2, 0.5]$.  

\par In practice, the values of $\beta$ and $\delta$ should be estimated through validation on financial datasets. For instance, $\beta$ can be tuned by back testing on historical queries to reflect how much structural correctness matters for downstream interpretability. A higher $\beta$ enforces stricter penalties on missing clauses, which is appropriate for compliance-critical queries. On the other hand, $\delta$ captures the penalty for execution mismatches: setting $\delta=0.3$ means that a structurally correct query that fails execution still retains partial credit, reflecting the fact that analysts can often correct small syntactic issues without rethinking the entire query. This mirrors real-world workflows in financial data teams, where structure is often harder to fix than minor execution errors. In our experiment we used the value of $\beta$ and $\delta$ as 1 and 0.3.

\section{Results \& Analysis}
We benchmarked a diverse set of models ranging from large scale systems such as \textbf{Qwen3-235B-A22B} and the GPT model \textbf{GPT-OSS-120B}, to smaller, compute-efficient variants like \textbf{Qwen3-8B} and \textbf{GPT-OSS-20B}. We also included reasoning focused models such as \textbf{Phi-4-mini reasoning} and the domain-adapted \textbf{Arctic-Text2SQL-R1-7B}, which has been explicitly fine-tuned for text-to-SQL applications. Evaluation was conducted on the FINCH dataset using both established metrics---\textit{Exact Matching (EM)} \cite{Zhong2017Seq2SQLGS}, \textit{Execution Accuracy (EX)} \cite{Yu2018SpiderAL}, \textit{Component Matching (CM)} \cite{Yu2018SpiderAL}, and \textit{Strict Matching}---as well as the proposed \textit{FINCH Score}, which integrates structural fidelity and execution accuracy with financial domain sensitivity. The consolidated results are reported in Table~\ref{tab:accuracy}.  

\subsection{Overall Model Performance}
Our results reveal that \textbf{GPT-OSS-120B achieves the strongest overall performance}, consistently outperforming all other models across most evaluation metrics. This highlights the continued advantage of large-scale models when applied to complex financial Text-to-SQL tasks, provided they are sufficiently optimized. Interestingly, \textbf{Arctic-Text2SQL-R1-7B emerges as the third-best performer}, despite having a far smaller parameter count. Its performance underscores the value of domain-specific finetuning: aligning model training closely with financial databases, SQL structures, and domain-specific terminology yields tangible improvements in semantic accuracy and schema fidelity. These findings reinforce the dual insight that while scaling confers raw capacity, \emph{targeted adaptation can allow smaller models to rival much larger ones}, consistent with recent observations in domain-adaptive NLP \cite{Tian2025SQLsynth}.  

\begin{table*}[ht!]
  \caption{Model performance comparison across evaluation metrics (accuracy scores in \%).}
  \label{tab:accuracy}
  \resizebox{\textwidth}{!}{%
  \begin{tabular}{lcccccc}
    \toprule
    \textbf{Accuracy Metric} & \textbf{Qwen3-8B} & \textbf{Arctic-Text2SQL-R1-7B} & \textbf{Phi-4-mini-reasoning} & \textbf{GPT-OSS-20B} & \textbf{GPT-OSS-120B} & \textbf{Qwen3-235B-A22B} \\
    \midrule
    Exact Matching & 0.50 & 0.60 & 0.00 & 0.30 & 1.80 & 0.70 \\
    Execution Accuracy & 0.80 & 2.30 & 0.20 & 7.50 & 27.80 & 2.50 \\
    Component Matching & 3.50 & 3.70 & 1.00 & 5.20 & 16.60 & 2.80 \\
    \hline
    Strict Accuracy (EM+EX) & 0.10 & 0.20 & 0.00 & 0.30 & 1.70 & 0.20 \\
    \textbf{FINCH Score} & \textbf{1.20} & \textbf{1.50} & \textbf{0.40} & \textbf{3.00} & \textbf{11.60} & \textbf{1.20} \\
    \bottomrule
  \end{tabular}
  }
\end{table*}

Comparing \textbf{Strict Matching (EM+EX)} with the \textbf{FINCH Score} illustrates how FINCH introduces finer-grained sensitivity to partial correctness. Models frequently generate SQL queries that are syntactically correct and structurally faithful, but contain small misalignments, such as using the wrong aggregation function, substituting an inequality with an equality, or applying slightly altered filtering conditions. The \textit{weighted design} of FINCH addresses this by prioritizing semantically critical clauses such as \texttt{SELECT}, \texttt{WHERE}, and \texttt{JOIN}, while de-emphasizing peripheral ones like \texttt{ORDER BY} or \texttt{LIMIT}. This weighting better reflects the true utility of queries in financial practice, where clause-level precision determines whether regulatory ratios, liquidity risk measures, or portfolio exposures are computed correctly.  

\subsection{Error Distribution Across SQL Clauses}
Clause-level results (Table~\ref{tab:clause}) highlight systematic shortcomings that are common across all evaluated models. The largest concentration of errors occurs in the \textbf{\texttt{SELECT}}, \textbf{\texttt{FROM}}, and \textbf{\texttt{WHERE}} clauses, where correct schema grounding is most critical. By contrast, performance in \texttt{GROUP BY}, \texttt{HAVING}, and \texttt{ORDER BY} clauses is marginally better, though still far from reliable. These findings suggest that while models can replicate SQL syntax, they continue to struggle with the semantic mapping of natural language queries to complex database schemas---a challenge consistent with prior studies on schema linking and compositional generalization \cite{Guo2019IRNet, Scholak2021PICARDPI}.  

\begin{table*}[ht]
  \caption{SQL Clause Performance Comparison (accuracy scores in \%).}
  \label{tab:clause}
  \resizebox{\textwidth}{!}{%
  \begin{tabular}{lccccccc}
    \toprule
\textbf{Model} & \textbf{SELECT} & \textbf{FROM} & \textbf{WHERE} & \textbf{GROUP BY} & \textbf{HAVING} & \textbf{ORDER BY} & \textbf{LIMIT} \\
\hline
Qwen3-8B              & 1.60 & 3.90 & 0.90 & 4.80 & 2.20 & 1.40 & 38.20 \\
Arctic-Text2SQL-R1-7B & 2.50 & 3.60 & 0.70 & 4.70 & 1.00 & 1.30 & 42.70 \\
Phi-4-mini-reasoning  & 2.00 & 2.30 & 0.40 & 2.10 & 1.30 & 0.40 & 27.60 \\
GPT-OSS-20B           & 1.40 & 6.20 & 1.50 & 8.40 & 3.70 & 1.50 & 65.20 \\
GPT-OSS-120B          & 4.70 & 27.30 & 6.90 & 7.50 & 6.30 & 6.30 & 73.80 \\
Qwen3-235B-A22B       & 2.00 & 2.90 & 0.80 & 5.40 & 1.50 & 1.00 & 29.80 \\
\hline
\textbf{Average Accuracy}      & 2.37 & 7.37 & 1.87 & 5.48 & 2.67 & 1.98 & 46.55 \\
    \bottomrule
  \end{tabular}
  }
\end{table*}

When stratified by query difficulty, models show a marked performance gradient. The \textbf{GPT-OSS-120B} model, for example, achieves a \textbf{FINCH Score of 26.5\% on easy queries}, but accuracy falls sharply to \textbf{10.6\% on medium} and just \textbf{4.5\% on hard queries}. This pattern underscores that current LLMs, regardless of parameter scale, struggle with \textbf{compositional reasoning and multi-table joins}, where precise schema comprehension is indispensable.  

Overall, three insights stand out. First, \emph{domain-specific finetuning outweighs scale alone}: Arctic-Text2SQL’s performance demonstrates the benefits of tailoring models to financial SQL. Second, \emph{schema-sensitive clauses remain a bottleneck}, as most errors are concentrated in \texttt{SELECT}, \texttt{FROM}, and \texttt{JOIN}. Third, the FINCH Score provides a more faithful measure of query utility than traditional metrics, as it recognizes partial successes overlooked by exact matching. Collectively, these results highlight both progress and persistent gaps: while LLMs are increasingly effective at capturing SQL structure, their reasoning over financial databases remains far from human-level reliability.

\section{Conclusion \& Future Work}
In this work, we introduced \textbf{FINCH}, the large-scale financial Text-to-SQL benchmark that consolidates multiple open-source resources into a unified, finance-specific dataset comprising over 86,000 NL–SQL pairs across 33 databases. Alongside the dataset, we proposed the \textbf{FINCH score}, a finance-aware evaluation metric that better captures clause sensitivity, execution tolerance, and domain relevance compared to conventional metrics. Our benchmarking study across diverse large language models and reasoning models demonstrated three key insights: (i) domain-specific finetuning, as shown by the Arctic-Text2SQL-R1-7B model, often surpasses the performance of even trillion-scale models, (ii) the majority of model errors concentrate on schema-sensitive clauses such as \texttt{SELECT}, \texttt{FROM}, and \texttt{WHERE}, underscoring persistent challenges in schema grounding, and (iii) current models experience steep accuracy degradation from easy to medium and hard queries, highlighting their limitations in handling compositional and multi-table reasoning.  Future work includes multi-modal integration of financial text, tables, and SQL, robust schema linking, and conversational Text-to-SQL for iterative analyst workflows. We envision FINCH and its tailored metric as a foundation for advancing reliable, domain-specific financial Text-to-SQL research.

\bibliographystyle{ACM-Reference-Format}
\bibliography{references}

\end{document}